# Applying Conditional Generative Adversarial Networks for Imaging Diagnosis

Haowei Yang[1], Yuxiang Hu[2], Shuyao He[3], Ting Xu[4], Jiajie Yuan[5], Xingxin Gu[6]

[1]University of Houston, Houston, USA

[2]Johns Hopkins University, Baltimore, USA

[3]Northeastern University, Boston, USA

[4]University of Massachusetts Boston, Boston, USA

[5]Brandeis University, Waltham, USA

[6]Northeastern University, Boston, USA

*Abstract*—This study introduces an innovative application of Conditional Generative Adversarial Networks (C-GAN) integrated with Stacked Hourglass Networks (SHGN) aimed at enhancing image segmentation, particularly in the challenging environment of medical imaging. We address the problem of overfitting, common in deep learning models applied to complex imaging datasets, by augmenting data through rotation and scaling. A hybrid loss function combining L1 and L2 reconstruction losses, enriched with adversarial training, is introduced to refine segmentation processes in intravascular ultrasound (IVUS) imaging. Our approach is unique in its capacity to accurately delineate distinct regions within medical images, such as tissue boundaries and vascular structures, without extensive reliance on domain-specific knowledge. The algorithm was evaluated using a standard medical image library, showing superior performance metrics compared to existing methods, thereby demonstrating its potential in enhancing automated medical diagnostics through deep learning

## I. Introduction

Intravascular ultrasound (IVUS) is an important method in the diagnosis and intervention of coronary artery disease. The extraction and identification of the boundary between the endovascular intima (LU) and the meso-outer membrane (MA) is the key to accurately evaluate the difficult lesions of coronary artery imaging and to intervene in coronary heart disease. In practical application, endovascular ultrasound imaging of the intima and the meso-outer membrane has a rich theoretical basis and rich clinical experience for doctors. The critical edge of intravascular ultrasound images is subjective and easy to be affected by artifacts and other perturbations. The analysis of images with several to one hundred frames is time-consuming and difficult to reproduce. The subjective judgment errors between and within physicians can be effectively eliminated, the image artifacts and disturbance effects can be reduced, and the time of diagnosis and treatment can be shortened. It is of great theoretical and practical value for the diagnosis and treatment of congenital heart disease to study the method of detecting the edge of the middle and outer membrane of the coronary artery.

Most existing algorithms can be divided into three aspects. The first type is based on direct detection and mining low-level features[1-2], including graph search[2-3], active contour[4-6], and graph-cutting algorithms [7-9]. This method has a great effect on calcium plaques in intravascular ultrasound images. The second is based on statistics and probability, using the optimal energy equation and the probability of the region of the largest possible pixel to obtain the optimal edge geometry. However, the influence of intravascular ultrasound shadow, vascular bifurcation and fibrous plaque often leads to the local extreme value of its evolution curve, resulting in huge protrusions or pits between the intima or meso-outer membrane [10]. The third stage is based on machine learning and consists of two stages: hand-set feature extraction and classifier optimization[11-14]. Among them, support vector machine[15], AdaBoost, error correction output code, and shallow ANN based on manual characteristics are important components of such algorithms. Although there have been many research results, the recognition rate is easily affected by external disturbance because only low-level features designed by humans are used. In addition, due to the complexity of feature extraction, the generalization ability and applicability of this method are not high[16]. In this paper, C-ivus GAN-SHGN, an algorithm based on the layered hourglass structure[17], was constructed and used to extract the edges of the inner and outer membranes of IVUS vessels. First of all, the intravascular ultrasound images were efficiently augmented to increase the sample size of intravascular ultrasound images. Three-dimensional reconstruction of IVUS was performed by C-ivus GAN-SHGN, and the IVUS was divided into three regions: surrounding tissue, plaque and lumen.

## II. RELEVANCE THEORY

The network can be regarded as a mapping between the random noise Z and the output screen[18]. The algorithm generates data by introducing a random noise C in the generation process, and then sends this process to the discriminator[19]. The results are then fed back to the synthesizer.

$$\min_F \max_S Q(S,F) = W_{u\sim Gdata(u)}[\log(S(u))] \\ + W_{u\sim G_c(c)}[\log(1-S(F(c)))] \quad (1)$$

$G_{data(u)}$ is the actual data distribution. $G_{c(c)}$ is the generated data distribution.

### A. Medical image enhancement based on CGAN

#### 1) CGAN

CGAN adds additional label information to production and discriminative networks, and uses additional constraints to guide the generation of samples [20]. This project intends to use the medical image with random noise C as the constraint information, and train the original image to make it correspond to the real image, so as to realize the effective enhancement of the image. The objective function Q (S, F) of CGAN is as follows:

$$\min_F \max_S Q(S,F) = \\ W_{u\sim Gdata(u)}[\log(S(u,v))] + \\ W_{u\sim G_c(c)}[\log(1-S(u,F(u,c)))] \quad (2)$$

The function S (u, v) is the possibility that the image input to the discriminator S is true, and the image $S(u,F(u,c))$ input to the discriminator S is generated by the generator F.

#### 2) Signal generator

In this paper, the CGAN architecture is mainly used, and U-Net network is used instead of CNN to retain the original image in the image to the maximum extent. Its construction is similar to the codec, but it uses the skip connection technology[21]. This structure is shown in Figure 1. This method makes full use of the bottom and top feature maps, which can better preserve the details of the image.

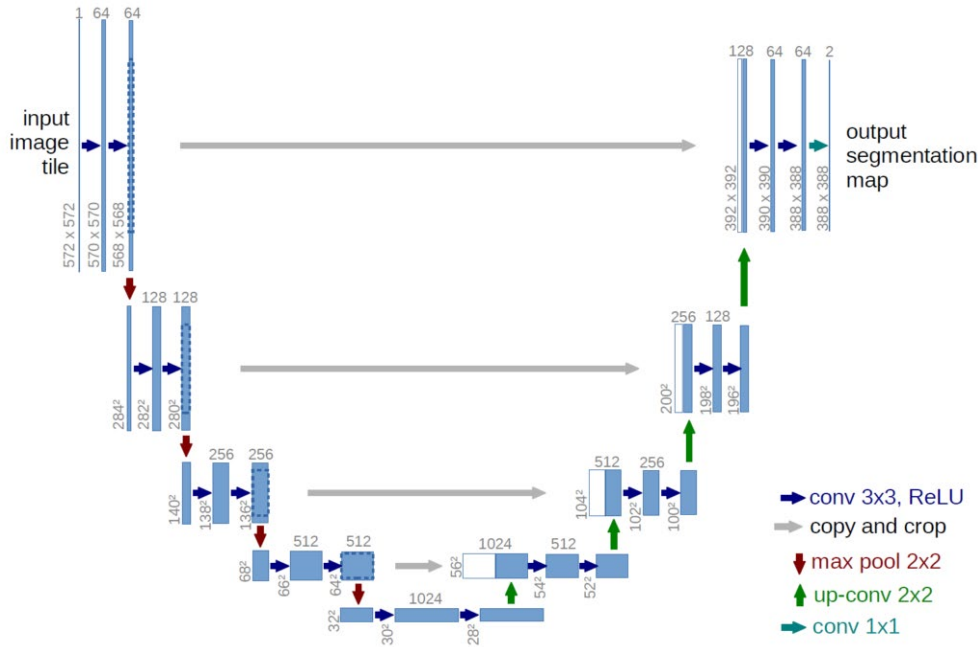

Fig. 1. U-Net and its compilation/decoding architecture

A network consisting of 8 convolutional networks is established. The first component of the generator is an encoder that extracts the features of the input medical image by doing a convolution drop sampling operation, and then uses the batch normalization (BN) method and the Leaky ReLU function as the excitation function. By using deconvolution neural networks, the low-dimensional features are reduced to the initial dimension to achieve accurate medical images. In order to avoid over-learning, this paper uses BN and Dropout operations to deal with the features of layer unwrapping, and then uses Relu functionality to achieve network activation and combines it with Tanh functionality. The network mode of the paper generator F is shown in Figure 2.

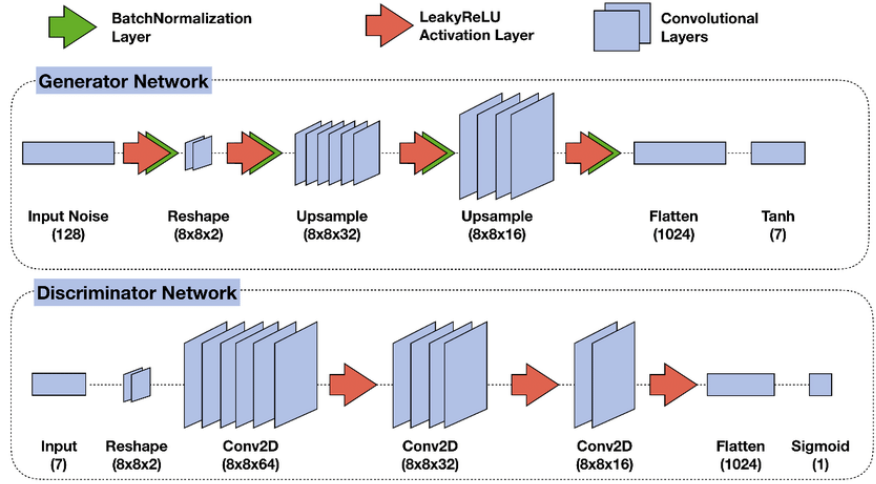

Fig. 2. Generator network structure

*3) Discriminator*

The resulting medical image with a clear image is used as an input to the discriminator S. All convolutional networks use the BN and Leaky ReLU activation layers. The Sigmod algorithm is used to optimize the network. The identification results correspond to the range of [0,1]. Result 1 indicates the sharpness of the medical image. 0 represents the medical image produced by generator F. The network pattern of the discriminator S is shown in Figure 3.

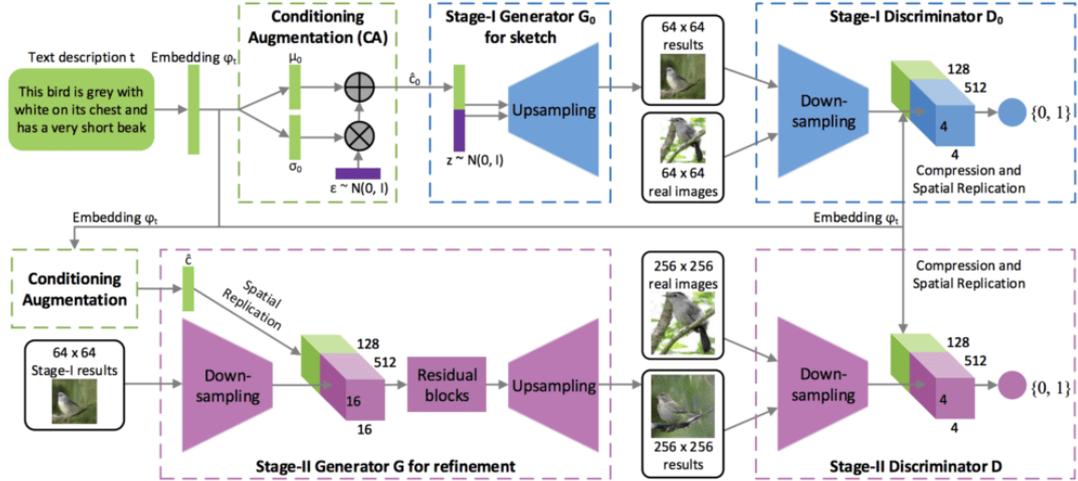

Fig. 3. Discriminator network structure

### B. Loss function

In the test model, CGAN loss function is introduced and $Loss_1$ loss function is added to ensure the similarity of the two images, considering that there is a large amount of data sharing between the two systems. In a standard CGAN network, the loss function is as follows

$$Loss_{CGAN}(F,S) = W_{u,v}[\log S(u,v)] + W_{u,c}[\log(1-S(u,F(u,c)))] \quad (3)$$

Where F is the generator, S is the discriminator, u is the input medical image, and V is the actual image. To get more low-frequency information and improve the sharpness of the resulting image, add penalty $Loss_1$ to F,

$$Loss_{Loss_1}(F) = W_{u,v,c}[\|v - F(U,C)\|] \quad (4)$$

By introducing the super parameter $\eta$ to achieve the balance of the two losses, it not only retains the characteristics of the high frequency band of the image, but also retains part of the low frequency information, so as to obtain a more accurate imaging effect. The loss function used in the final paper is as follows

$$Loss = \arg\min_F \max_S Loss_{CGAN}(F,S) + \eta Loss_{Loss_1} \quad (5)$$

### III. SYSTEM WORKFLOW

An adversarial modeling method based on Civus GAN-SHGN was used to divide the segmented IVUS images into three categories: surrounding tissue, plaque and lumen. Then the Civus GAN-SHGN image was binarized to obtain the boundary between the middle-outer membrane and the inner-outer membrane.

## A. C-ivusGAN-SHGN network framework

The former is a user-specified probability density function of the image, with an explicit distribution. The latter can be used for image data without explicit function assignment. The generator is designed to produce real, authentic biological or medical image data, which contains the segmentation of human organs.

## B. Training of C-ivusGAN-SHGN

The above research work is carried out based on the TensorFlow open-source development platform. The negative slope of LU activation in encoder, decoder and discriminator groups is 0.2.

## IV. RESULTS

### A. Comprehensive loss function

The comprehensive loss function of C-ivus GAN-SHGN is divided into two parts: reconstruction loss and resistance loss. The L1 and L2 spacing is used as the cost of resisting network reconstruction so that the resulting image is consistent with the single contour obtained by the physician. Only L1 or L2 reconstruction loss can be used to obtain the optimal score. The concept of adversarial learning is introduced into the algorithm, which makes the generated image have better segmentation effect. The reason is that the spatial distribution between the images obtained by the anti-loss algorithm and the contours of multiple patients on the same sample is basically the same (Table 1).

TABLE I. ANALYSIS OF COMMON LOSS FUNCTION OF GENERATIVE ADVERSARIAL NETWORK

| Evaluation index | Counter loss | L1 loss | L2 loss | Confrontations and L1 losses | Confrontation and L2 loss |
|---|---|---|---|---|---|
| LU-JM | 0.8968 | 0.9010 | 0.9182 | 0.9177 | 0.9206 |
| MA-JM | 0.9185 | 0.9203 | 0.9217 | 0.9290 | 0.9223 |
| LU-PAD/% | 4.8114 | 4.4253 | 3.9897 | 3.9204 | 3.3066 |
| MA-PAD/% | 5.4945 | 5.7321 | 5.7321 | 4.8411 | 10.7712 |
| LU-HD/mm | 0.2813 | 0.2120 | 0.2026 | 0.2093 | 0.2020 |
| MA-HD/mm | 0.2761 | 0.4023 | 0.2177 | 0.2166 | 0.2258 |
| LU-AD/mm | 0.0816 | 0.0764 | 0.0586 | 0.0634 | 0.0566 |
| MA-AD/mm | 0.0870 | 0.0662 | 0.0725 | 0.0609 | 0.0599 |

Since lumen area and lumen area can be regarded as two regions with strong consistency, they can be classified as one region. The L2 loss method was used to divide the A section twice, and a relatively high score of each evaluation index was obtained. The speckle area and the gray scale of zone A vary very unevenly (Figure 4), and there are obvious mutations in the two intervals. If there are calcification spots in the spots, its gray change will be more drastic.

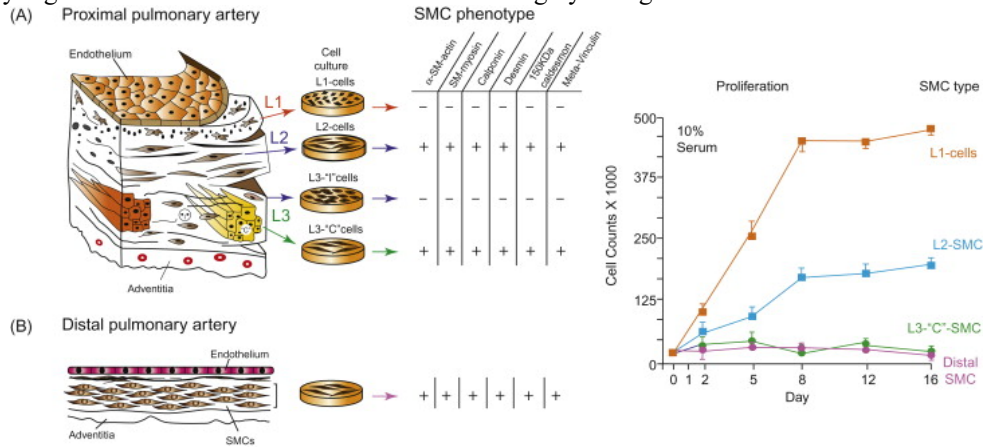

Fig. 4. Profile line analysis of the inner and middle - outer membranes

As can be seen from Table 2, when the super parameter a=1 to 128. When b=32, C-ivus GAN-SHGN can better extract the edge of vascular intima in IVUS images. L2 is used as a reconstruction loss for the super parameter a=1, as shown in Table 3. Civus GAN-SHGN is in the IVUS image. In the case of b=64, the intimal edge can be extracted well. In the case of b=128, C-ivus GAN-SHGN can better extract the boundary of the middle - outer membrane of IVUS. It can be seen from Table 2 and Table 3 that the super parameter b has no obvious influence on the image segmentation effect. Better edge detection results can be obtained in the range of 32 to 128. The following tests will be uniformly set to b=100.

TABLE II. RECONSTRUCTION LOSS L1

| | β | | | | | | | |
|---|---|---|---|---|---|---|---|---|
| | 1 | 2 | 4 | 8 | 16 | 32 | 64 | 128 |
| LU-JM | 0.8811 | 0.8910 | 0.8910 | 0.9009 | 0.9108 | 0.9108 | 0.9108 | 0.9108 |

| | MA-JM | 0.9108 | 0.9108 | 0.9207 | 0.9207 | 0.9207 | 0.9207 | 0.9306 | 0.9306 |

TABLE III. RECONSTRUCTION LOSS $L_2$

| | β | | | | | | | |
|---|---|---|---|---|---|---|---|---|
| | 1 | 2 | 4 | 8 | 16 | 32 | 64 | 128 |
| LU-JM | 0.8910 | 0.9207 | 0.9207 | 0.9108 | 0.9108 | 0.9207 | 0.9207 | 0.9207 |
| MA-JM | 0.9108 | 0.9207 | 0.9207 | 0.9207 | 0.9207 | 0.9207 | 0.9207 | 0.9207 |

## B. Impact of generator network structure

FCN, U-Net, SegNet, Deconv Net, etc. are typical image partitioning networks with semantic features, and they can all be used to construct C-ivus GAN-SHGN. This paper discusses the influence of three construction methods on image cutting, namely Pix2Pix-1 based on U-Net structure and codec structure. During the refactoring process, the user can reuse the low-level features inside the encoder to accurately restore the details of the original image. The main feature of the system is the combination of encoder and decoder (Figure 5 cited in Appl.Sci.2022, 12(1), 403).

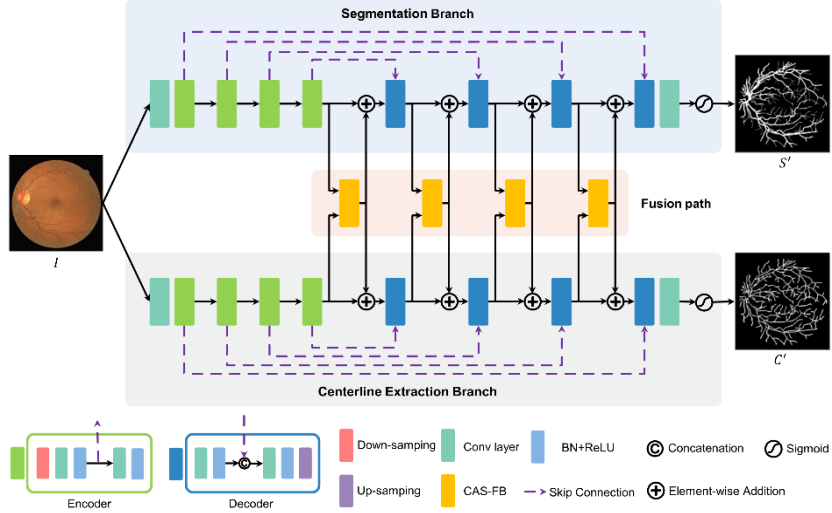

Fig. 5. Three different generator network structures in C-ivusGAN-SHGN

The comparison in Table 4 shows that in C-ivus GAN-SHGN algorithm, encoder decoder is slightly inferior in image segmentation effect compared with Pix2Pix-1 algorithm based on U-Net architecture. However, the hourglass SHGN based on the middle layer structure has more advantages than Pix2Pix-1. The stacked hourglass has a tighter mesh structure and is smaller than the Pix2Pix-1.

TABLE IV. EFFECTS OF DIFFERENT GENERATOR NETWORK STRUCTURES

| Evaluation index | Pix2Pix-1(U-Net) | Pix2Pix-2(E-D) | Method 1(no inputs) | Method 2(within puts) |
|---|---|---|---|---|
| LU-JM | 0.9128 | 0.9045 | 0.9090 | 0.9177 |
| MA-JM | 0.9279 | 0.9195 | 0.9196 | 0.9290 |
| Model size /M | 226.4130 | 79.7940 | 158.6970 | 158.6970 |

## V. CONCLUSION

This paper constructs an algorithm based on a hierarchical structure for modeling research. First, the adversarial training and CGAN methods were used to take the ultrasonic image as the constraint and corresponding it to the segmented image, so as to realize the three main regions of the IVUS image: surrounding tissue, plaque and lumen. According to the image segmentation results, the edge between the middle - outer membrane and the inner - outer membrane is extracted by threshold analysis method. C-ivus GAN-SHGN takes layered hourglass neural network as its basic structure, which is compact and has fewer parameters, and is superior to Pix2Pix UNet. Our results, indicating a marked improvement over existing methods, confirm the robustness and applicability of our model. Looking forward, the integration of C-GAN with SHGN presents a promising avenue for further research in other areas of medical imaging and could potentially be adapted for real-time diagnostic systems. The adaptability of our model to different types of medical imaging modalities also suggests extensive possibilities for future interdisciplinary research, aimed at bridging the gap between advanced machine learning techniques and clinical applications.